\documentclass[nohyperref]{article}

\usepackage{microtype}
\usepackage{graphicx}
\usepackage{subfigure}

\usepackage{booktabs} 

\usepackage{hyperref}

 \usepackage[accepted]{icml2022}

\usepackage{amsmath}
\usepackage{amssymb}
\usepackage{mathtools}
\usepackage{amsthm}

\usepackage[capitalize,noabbrev]{cleveref}

\theoremstyle{plain}

\theoremstyle{definition}

\theoremstyle{remark}

\usepackage[textsize=tiny]{todonotes}

\icmltitlerunning{Cosmological constraints using simulation based inference}

\begin{document}

\twocolumn[
\icmltitle{Estimating Cosmological Constraints from Galaxy Cluster Abundance using Simulation-Based Inference}



\icmlsetsymbol{equal}{*}

\begin{icmlauthorlist}
\icmlauthor{Moonzarin Reza}{equal,xxx,yyy}
\icmlauthor{Yuanyuan Zhang}{equal,xxx,yyy}
\icmlauthor{Brian Nord}{comp,abc,def}
\icmlauthor{Jason Poh}{comp,abc}
\icmlauthor{Aleksandra Ciprijanovic}{def}
\icmlauthor{Louis Strigari}{xxx,yyy}
\end{icmlauthorlist}

\icmlaffiliation{xxx}{
 Department of Physics and Astronomy, Texas A\&M University, College Station, TX 77843, USA}

\icmlaffiliation{yyy}{
Mitchell Institute for Fundamental Physics and Astronomy, Texas A\&M University, College Station, TX 77843, USA}

\icmlaffiliation{comp}{Department of Astronomy and Astrophyiscs, University of Chicago, Chicago, IL 60637, USA}
\icmlaffiliation{abc}{Kavli Institute for Cosmological Physics, University of Chicago, Chicago, IL 60637, USA}
\icmlaffiliation{def}{Fermi National Accelerator Laboratory, Batavia, IL 60510, USA}

\icmlcorrespondingauthor{Moonzarin Reza}{moonzarin@tamu.edu}

\icmlkeywords{Cosmological constraints, simulation-based inference, machine learning}

\vskip 0.3in
]



\printAffiliationsAndNotice{\icmlEqualContribution} 

\begin{abstract}
 
Inferring the values and uncertainties of cosmological parameters in a cosmology model is of paramount importance for modern cosmic observations.
In this paper, we use the simulation-based inference (SBI)  approach to estimate cosmological constraints from a simplified galaxy cluster observation analysis. Using data generated from the Quijote simulation suite  and analytical models, we train a machine learning algorithm to learn the probability function between cosmological parameters and the possible galaxy cluster observables. The posterior distribution of the cosmological parameters at a given observation is then obtained by sampling the predictions from the trained algorithm. Our results show that the SBI method can successfully recover the truth values of the cosmological parameters within the 2$\sigma$ limit for this simplified galaxy cluster analysis, and acquires similar posterior constraints obtained with a likelihood-based Markov Chain Monte Carlo method, the current state-of the-art method used in similar cosmological studies. 
\end{abstract}

\section{Introduction} \label{introduction}


Studying cosmic structure formation and matter clustering statistics \citep[e.g.,][]{2008ARA&A..46..385F,2013PhR...530...87W, 2022PhRvD.105b3520A} in the late Universe (after redshift 2.0) has provided a rich ground for constraining cosmology models. These studies include analyses of galaxy clusters \citep[see reviews in ][]{2011ARA&A..49..409A, 2012ARA&A..50..353K}, the largest gravitationally-bound structures in the Universe. The abundances and weak lensing mass measurements of galaxy clusters have been used to provide competitive constraints for $\mathrm{\Lambda CDM}$ models in ongoing cosmic surveys like the Dark Energy Survey (DES) \citep{2020PhRvD.102b3509A, 2021PhRvL.126n1301T} , and is also projected to be powerful for studying $\mathrm{wCDM}$ models in furture experiment like the Legacy Survey of Space and Time (LSST) \citep{2018arXiv180901669T} at the Vera C. Rubin Observatory. 

Moving forward, one potential challenge with those cosmic structure studies is about efficiently acquiring Bayesian posterior constraints of cosmological parameters, with the increasingly complicated cosmological and astrophysical models, and the larger parameter space generated by those models. To date, many of the cosmic structure formation studies rely on a Markov Chain Monte Carlo (MCMC) method to sample parameter posterior distributions, which further requires calculating a likelihood at running time based on theoretical models for the observables, and  takes an increasingly long computing time which is no longer convenient \citep{2022arXiv220208233L}. 

A potential improvement to those analyses is to adopt Machine Learning through a so-called ``Simulation-Based Inference" \citep[see a review in ][]{cranmer2020frontier}, sometimes also known as ``likelihood-free" approach \citep[e.g., see ][ for a closely related application]{2022ApJ...925..145T}. In this approach, we may precompute mock observables based on the theoretical models, known as ``simulations", and then use those ``simulations" to train a machine learning based inferer to map out the probabilistic function (or the ``likelihood" function in alternative set-ups) between the model parameters and their possible observables. This probabilistic function can then be used to derive the posterior probability distribution of the model parameters at a given observable. 

In this note, we demonstrate the potential of this SBI approach by applying it to a simplified galaxy cluster abundance analysis described in Section~\ref{sec:2}. In Section~\ref{sec:3}, we describe the SBI method, the Quijote simultaions and the analytical models. We present our results and conclusions in Sections~\ref{sec:4}  and~\ref{sec:5} respectively.

\section{The Galaxy Cluster Analysis} \label{sec:2}

\label{sec:3}
\begin{figure}[ht]
\begin{center}
\centerline{\includegraphics[scale=1]
{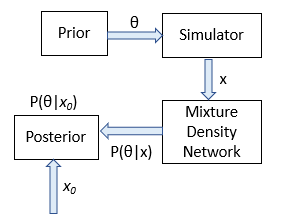}}
\caption{Block diagram illustrating the simulation-based inference method applied in this analysis.}
\label{fig:blocksbi1}
\end{center}
\end{figure}


Galaxy clusters correspond to the most massive (typical mass larger than $10^{14}\mathrm{M}_\odot/h$), gravitationally-bound structures in the universe, also known as dark matter halos in cosmic simulations. The abundance and mass distribution of massive dark matter halos are sensitive to cosmology, as indicated by the mass function of dark matter halos \citep[e.g.,][]{1974ApJ...187..425P, 2008ApJ...688..709T}. Observationally, galaxy cluster cosmology studies  often rely on the number counts of galaxy clusters and their average masses in an observational range as the observational data vectors, which can be considered as summary statistics of their observations. 

To resemble a typical galaxy cluster cosmology analysis, in this note, we analyze the dark matter halo counts and their masses in the Quijote simulations  \cite{villaescusa2020quijote} with a fiducial Planck cosmology model. These simulations use a box volume of ${(1 Gpc/h)}^3$, and follow the evolution of $256^3$ or more dark matter particles. Our cosmological observables consist of the mean masses and the number counts of galaxy cluster-sized dark matter halos in four mass bins $[10^{14.0}, 10^{14.2}]$, $[10^{14.2}, 10^{14.4}]$, $[10^{14.4}, 10^{14.6}]$ and $[10^{14.6}, +\infty)\mathrm{M}_\odot/h$ at two different values of redshifts (0 and 0.5).

We use those observables to constrain five cosmological parameters: matter density ($\Omega_m$), baryonic density ($\Omega_b$), Hubble’s constant ($h$), power law index of density perturbation ($n_s$), and the amplitude fluctuation of matter power spectrum ($\sigma_8$).
The fiducial cosmology parameters are set at $\Omega_m$ = 0.3175, $\Omega_b$ = 0.049, $h$  = 0.6711, $n_s$ = 0.9624 and $\sigma_8$ = 0.834 \cite{2020JCAP...03..040H}.  We use the dark matter halos in this  fixed-cosmology simulation suite as our observational test sample.

\section{SBI method, Quijote simulations and analytical models} \label{sec:3}

\begin{figure*}[ht]
\begin{center}
\centerline{\includegraphics[scale=0.5]{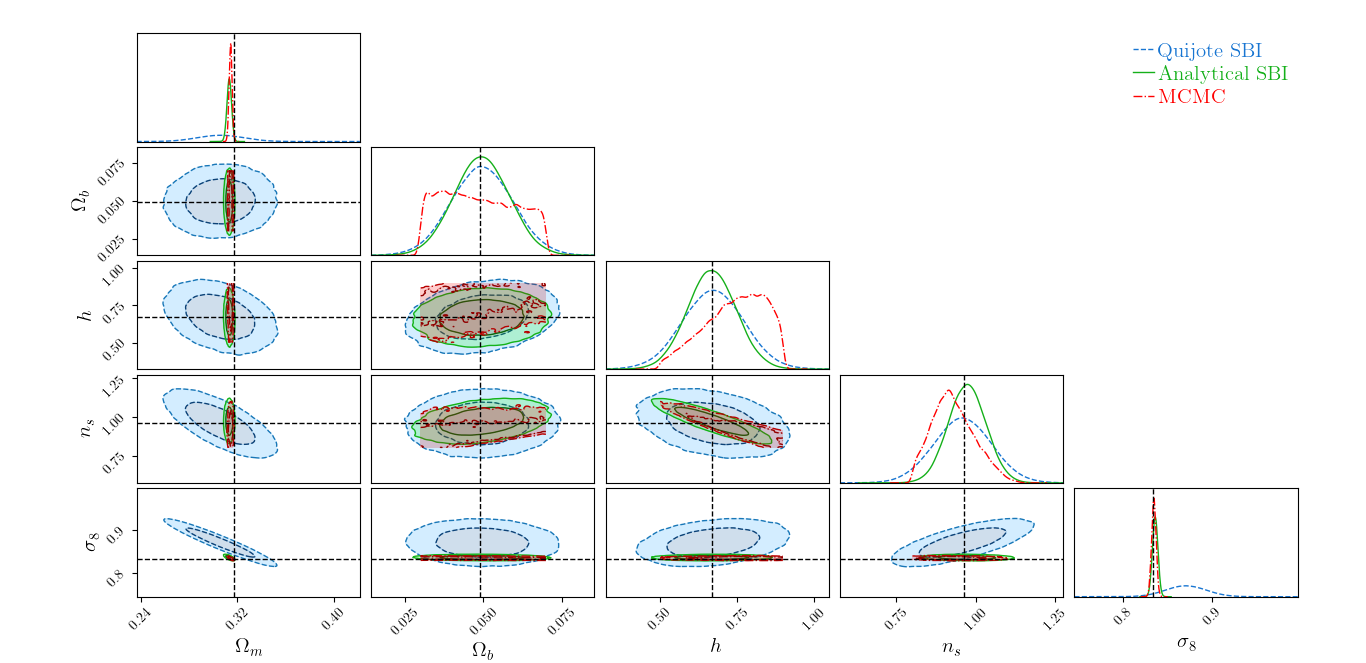}}
\caption{Posterior distributions of the cosmological parameters for analytical models (green solid), Quijote simulations (blue dashed) and MCMC (red dashdot)}
\label{fig:an+qui}
\end{center}
\end{figure*}

\subsection{Simulation-based inference (SBI) method}
Simulation-based inference (SBI) \cite{cranmer2020frontier}  methods can incorporate complex physical processes and observational effects in forward simulations \cite{2022ApJ...925..145T}. SBI is a machine-learning approach to learn the sampling distribution of data as a function of the model parameters. The main goal of simulation-based inference is to identify parameter sets which are both compatible with prior knowledge and match empirical observations. The outputs of SBI are not point estimates; rather all probabilistic values of parameter space consistent with the inputs are identified \cite{pmlr-v130-lueckmann21a}. We use \texttt{SBI} \cite{tejero2020sbi}, a neural network-based PyTorch package \cite{ketkar2021introduction}. The process is explained by the block diagram shown in Figure~\ref{fig:blocksbi1}. The prior is first sampled to obtain an initial set of parameters, which are used to create synthetic data using forward simulations. Data generated by forward simulations are then passed into a gaussian mixture density network. The output of this network is the probability distribution of the cosmological variables as a function of the observables (number counts and mean masses of dark matter halos). The posterior (probability distribution function) is then sampled to generate the parameter space of the cosmological variables for specific observables. One of the major advantages of SBI over traditional likelihood-based methods is that it can be implemented in stages -- simulations, training, inference. It provides us with an opportunity to more efficiently perform complicated analysis or analysis of large volume of data.

\subsection{Quijote latin-hypercube simulations}


 The Quijote simulation suite contains subsets of simulations which are run for different sets of cosmological parameters \citep{villaescusa2020quijote}. 
One such example is the latin-hypercube simulation set which contains 2000 simulations and their cosmological parameters are varied from the standard fiducial cosmology model  with $\Omega_m$ in the range of $[0.1 - 0.5]$, $\Omega_b = [0.03 - 0.07]$, $h = [0.5 - 0.9]$, $n_s = [0.8 - 1.2]$ and $\sigma_8 = [0.6 - 1.0]$. We use the same summary statistics (described in Section~\ref{sec:2}) for the galaxy cluster obervables from these simulations to train the machine learning model. 

\subsection{Analytical models} \label{sec:3.3}

The galaxy cluster observables can also be calculated using analytical formulas as: 

\begin{equation}
\begin{split}
N(\Delta_M|z,\Theta) & = V(z) \int_{\Delta_M} n(M|z, \Theta) + \delta_N,\\
NM(\Delta_M, z| \Theta) & = V(z) \int_{\Delta_M} M\times n(M |z,  \Theta), \\
M(\Delta_M, z| \Theta) & = NM(\Delta_M|z, \Theta)/N(\Delta_M|z, \Theta) + \delta_M.
\end{split}
\end{equation}

In these equations, $N(\Delta_M|z,\Theta)$ and $M(\Delta_M, z| \Theta)$ are the mock observables calculated from analytical models. $V(z)$ is the cosmic volume of the Quijote simulations that the problem is based upon. $n(M, z | \Theta)$ is the theoretical halo mass function that describes the number density of dark matter halos at redshift $z$, which has an analytical form depending on  cosmology, indicated by $\Theta$. In this analysis
$\Theta$ refers to the five cosmological parameters, $\Omega_m$, $\Omega_b$, $h$, $n_s$ and $\sigma_8$.
We make use of the \citet{2011ApJ...732..122B} halo mass function for the FOF halo mass definitions implemented in Colossus, with a linear correction \citep{2019MNRAS.488.4779C} to account for differences with the Quijote simulations. Furthermore, $\delta_N$ and $\delta_M$ represent noises added to the analytical models, caused by cosmic variance. We use the Quijote fixed cosmology simulations to estimate the cosmic variances for $N(\Delta_M|z,\Theta)$ and $M(\Delta_M|z,\Theta)$, and then randomly draw a gaussian uncertainty according to those variances, $\delta_N$ and $\delta_M$, for each set of mock observables. In the end, we generate those analytical model simulations for over 10,000 sets of cosmological parameters. Specifically, we use a total of  11378 simulations to train the model. It is to be noted that the fast analytical simulations give valid insights and  generating these simulations is feasible for this particular study.

 Later in this analysis, the  cosmic variances used in this fast analytical methods are  used as the  covariance matrices in a multi-dimensional Gaussian likelihood implemented in a Markov Chain Monte Carlo method as a comparison analysis. 

\section{Results} \label{sec:4}
\subsection{SBI method}

\begin{figure}[ht]
\begin{center}
\centerline{\includegraphics[width=\columnwidth]{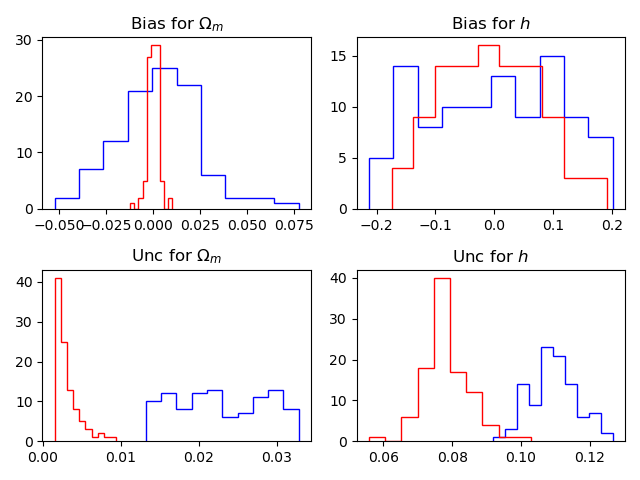}}
\caption{Distribution of bias (upper panel) and uncertainty (lower panel) for 100 test samples for  analytical models (red) and Quijote simulations (blue)}
\label{fig:newfig}
\end{center}
\end{figure}

\begin{figure}[ht]
\begin{center}
\centerline{\includegraphics[width=\columnwidth]{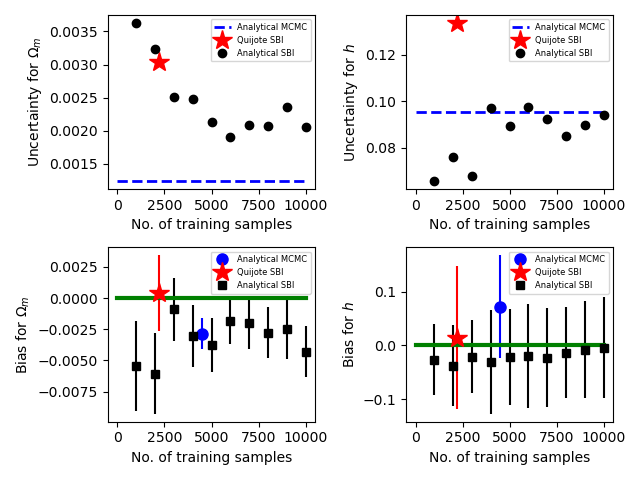}}
\caption{Uncertainty (upper panel) and bias (lower panel) for different number of training samples for  analytical models, Quijote simulations and MCMC}
\label{fig:mean+bias+ana}
\end{center}
\end{figure}

We derive the posterior distributions of the five cosmological parameters with SBI method using the Quijote simulations (blue dashed lines), and the analytical models (green solid lines) as the training samples. The black dashed lines represent the truth values of those parameters. For comparison, we show the posterior results from  Markov Chain Monte Carlo (MCMC) process, a state-of-the-art default method for sampling the posterior parameter distributions in such an analysis. In this comparison, the MCMC  results are shown by  red dashdot lines.

As shown in Figure~\ref{fig:an+qui} and listed in Table~\ref{table:q+aunc}, the truth values of the parameters are reasonably recovered by the MCMC method, and the SBI method applied to both the Quijote simulations and the analytical-model based simulations. 
For $\Omega_m$, the truth value is recovered  at 0.1$\sigma$, 1.6$\sigma$ and 2.2$\sigma$ level by the Quijote simulations, analytical models and MCMC method respectively. For the other four parameters, the truth values are recovered within the 1$\sigma$ range by all three methods. For $h$ and $n_s$, the SBI bias (both Quijote and analytical simulations) is much smaller than the corresponding MCMC bias.

In cosmological analysis, it is also important to accurately estimate the posterior uncertainties of the cosmological parameters, so as to correctly evaluate consistency and tensions between different cosmological models. While the MCMC method (state-of-the-art model) and the analytical-model based SBI method yield similar levels of uncertainties, the Quijote simulations based SBI method results in larger uncertainties than the other two methods for all parameters except $\Omega_b$. The histograms in Figure~\ref{fig:newfig} show the bias and uncertainty distributions for 100 test samples for the analytical simulations (red) and the Quijote simulations (blue) for $\Omega_m$ and $h$. Both methods result in symmetric bias distributions. For the analytical simulations, the bias tends to be distributed over a smaller range of values, and uncertainties are  smaller (same result as Figure~\ref{fig:an+qui}) than the Quijote simulations.  We suspect that this might be due to the training sample size being too small in the latin-hypercube (Quijote) based results, and explore further in the next subsection.


\begin{table}[t]
\caption{Estimates of cosmological parameters obtained using SBI and MCMC along with the truth values}
\label{table:q+aunc}

\begin{center}
\resizebox{\columnwidth}{!}{
\begin{tabular}{lcccr}
\toprule
 Parameters & Truth & Analytical & Quijote & MCMC \\
\midrule
$\Omega_m$ & 0.317    & $0.314^{+0.002}_{-0.002}$   & $0.317^{+0.003}_{-0.003}$ &  $0.315^{+0.001}_{-0.001}$ \\
\midrule
$\Omega_b$ & 0.049  & $0.051^{+0.009}_{-0.009}$& $0.050^{+0.011}_{-0.011}$ &  $0.049^{+0.011}_{-0.011}$\\
\midrule
$h$   &  0.671 &  $0.672^{+0.091}_{-0.091}$ & $0.679^{+0.119}_{-0.119}$&  $0.744^{+0.095}_{-0.095}$ \\
\midrule
$n_s$ & 0.962     &  $0.982^{+0.067}_{-0.067}$ &  $0.966^{+0.087}_{-0.087}$ &  $0.930^{+0.066}_{-0.066}$\\
\midrule
$\sigma_8$ & 0.834    & $0.836^{+0.003}_{-0.003}$ & $0.831^{+0.004}_{-0.004}$&  $0.835^{+0.003}_{-0.003}$\\
\bottomrule
\end{tabular}
}
\end{center}
\end{table}

 \subsection{Bias and uncertainty variation with  training sample size}
 
 We further evaluate how the bias and uncertainty level of the posterior constraints vary with different training sample sizes used in analytical-model based SBI method. We define bias and uncertainty ($\sigma$) as:

\begin{math}
    bias=\bar{\theta}_{Posterior}- \theta_{truth}
\end{math}

\begin{math}
 \sigma = \sqrt{\frac{ \sum_{n=1}^{N} (\theta_{Posterior, i}-\bar{\theta}_{Posterior})^2}{N}}
\end{math}

 In Figure~\ref{fig:mean+bias+ana}, we plot the  uncertainty (black circles) and bias (black squares) for different sizes of training sets using data generated from the analytical models for $\Omega_m$ and $h$ . We start with 1000 samples and increase the training sample size in steps of 1000 until we reach 10000 samples. Uncertainty and bias corresponding to the MCMC (blue dashed lines and blue circles) and latin-hypercube simulations (red stars) are also plotted on the same axes.  The uncertainty plots (upper panel) show that apart from some random fluctuations, uncertainty reduces with an increase in the number of training samples for $\Omega_m$ and stablizes when the sample size reaches $\sim$5000, for which the MCMC bias is about 33\% lower than the analytical-simulations SBI bias. A similar trend is also observed for $\sigma_8$ (not shown). There is no observed correlation between the training sample size and the uncertainty of $h$, $\Omega_b$ (not shown), and  $n_s$(not shown), indicating that the contraints are not affected by the training sample size we tested here. The convergent value of analytical SBI  uncertainty for these parameters is comparable to that of the MCMC method.

According to the results shown in the lower panel of Figure~\ref{fig:mean+bias+ana}, on average, the SBI bias based on analytical simulations diminishes as the size of training sample increases for $\Omega_m$. For $h$, the bias is nearly constant and independent of training sample size. Like the uncertainty, the bias for $\sigma_8$ shows a trend similar to $\Omega_m$, and the bias for $\Omega_b$ and $n_s$ follow the trend similar to h.   MCMC bias  is comparable to the SBI analytical bias for $\Omega_m$, and is higher than the  analytical bias for $h$.

\section{Conclusions} \label{sec:5}

We have applied a simulation-based inference method to a galaxy cluster cosmological analysis, using data generated from the Quijote latin-hypercube simulations and analytical models.  Our results show that SBI method can recover the truth cosmological parameters for this galaxy cluster analysis within the 2$\sigma$ limit. 
On average SBI method results in smaller bias than MCMC.  We have also evaluated the dependence of bias and uncertainty on the training sample size for the analytical simulations and conclude that the uncertainty converges for a sample size $\sim$5000.   The success of this attempt demonstrates that SBI is a promising method to be employed in future galaxy cluster cosmological analyses to shed light on the long-standing cosmological mysteries.



\bibliography{example_paper}
\bibliographystyle{icml2022}

\end{document}